\documentclass[journal= jctcce, manuscript=article]{achemso}

\usepackage[version=3]{mhchem} 
\usepackage{chemformula} 
\usepackage[T1]{fontenc} 

\usepackage{graphicx}
\usepackage{amsmath} 
\usepackage{mathtools}
\usepackage{amsfonts}
\usepackage{bbm}
\usepackage{braket}
\usepackage{caption}
\usepackage{graphicx}
\usepackage{float}
\usepackage{caption}
\usepackage{subcaption}
\usepackage{url}
\usepackage{color}
\usepackage[ruled]{algorithm2e}

\usepackage{amssymb}
\usepackage{cancel}
\usepackage{upgreek}
\usepackage{tabularx}
\usepackage{color}
\usepackage{units}
\usepackage{hhline}
\usepackage{textfit} 

\usepackage{slashed}
\graphicspath{ {images/} }

\usepackage{graphicx}
\usepackage{dcolumn}
\usepackage{bm}
\usepackage{hyperref}
\usepackage[mathlines]{lineno}

\usepackage{tikz}
\usetikzlibrary{positioning}
\usepackage{qcircuit}
\usepackage{braket}
\usepackage{indentfirst}
\usepackage{float}

\author{Kumar J. B. Ghosh}
\author{Sabre Kais}
\affiliation[Purdue University]{Department of Chemistry and Physics, Purdue University,\\West Lafayette, Indiana 47907, USA}
\author{Dudley R. Herschbach}
\email{dherschbach@gmail.com}
\affiliation[Harvard University]{Department of Chemistry and Chemical Biology, Harvard University \\Cambridge MA 02138, USA}
\title[interpolation] {Unorthodox dimensional interpolations for He, Li, Be atoms and hydrogen molecule}

\abbreviations{IR,NMR,UV}

\begin{document}


\begin{abstract}
We present a simple interpolation formula using dimensional limits $D=1$ and $D=\infty$ to obtain the $D=3$ ground-state energies of atoms and molecules. For atoms, these limits are linked by first-order perturbation terms of electron-electron interactions. This unorthodox approach is illustrated by ground-states for two, three, and four electron atoms, with modest effort to obtain fairly accurate results. Also, we treat the ground-state of H$_2$ over a wide range of the internuclear distance R, and compares well with the standard exact results from the Full Configuration Interaction method. Similar dimensional interpolations may be useful for complex many-body systems.   
\end{abstract}


\newpage

\section{\label{sec:Introduction} Introduction}

Dimensional scaling, as applied to chemical physics, offers promising computational strategies and heuristic perspectives to study electronic structures and obtain energies of atoms, molecules and extended systems \cite{Stillinger1975,MLODINOW1980314,yaffe1982large,Herschbach1986}. Taking a spatial dimension other than $D=3$ can make a problem much simpler and then use perturbation theory or other techniques to obtain an approximate result for $D=3$. Years ago, a D-scaling technique used with quantum chromodynamics \cite{witten1980} was prompted for helium \cite{MLODINOW1980314,yaffe1982large,Herschbach1986}. The approach began with the $D \to \infty$ limit and added terms in powers of $\delta=1/D$. It was arduous and asymptotic but by summation techniques attained very high accuracy for $D=3$ \cite{goodson1992large}. Other dimensional scaling approaches were extended to N-electron atoms \cite{Zhen1993}, renormalization with $1/Z$ expansions \cite{kais19941}, random walks \cite{rudnick1987shapes}, interpolation of hard sphere virial coefficients \cite{kais-hard-sphere}, resonance states \cite{kais-resonances} and dynamics of many-body systems in external fields \cite{kais-laser1,kais-laser2}.

Recently, a simple analytical interpolation formula emerged using both the $D=1$ and $D\to\infty$ limits for helium \cite{Herschbach2017}. It makes use of only the dimensional dependence of a hydrogen atom, together with the exactly known first-order perturbation terms with $\lambda = 1/Z$ for the dimensional limits of the electron-electron $\langle1/r_{12}\rangle$ interaction. In the $D=1$ limit, the Columbic potentials are replaced by delta functions in appropriately scaled coordinates \cite{Rosenthal1971}. In the $D\to\infty$ limit, the electrons assume positions fixed relative to another and to the nucleus, with wave functions replaced by delta functions \cite{Loeser1986}. Then at $D=3$, the ground state energy of helium $\epsilon_3$ can be obtained by linking $\epsilon_1$ and $\epsilon_\infty$ together with the first-order perturbation coefficients $\epsilon_1^{(1)}$ and $\epsilon_\infty^{(1)}$ of the $1/Z$ expansion. The first-order terms actually provide much of the dimension dependence. This article exhibits the applicability of an unorthodox formula, a blend of dimensions with first-order perturbations, to more complex many-body systems.

We outline the following sections: in \ref{sec:Dimensional interpolation formula} the interpolation formula; in \ref{sec:Dimensional scaling and interpolation for two-electron atoms} treat helium; in \ref{sec:Dimensional scaling and interpolation for three-electron atoms} lithium; in \ref{sec:Dimensional scaling and interpolation for four-electron atoms} beryllium; in \ref{sec:Dimensional scaling and interpolation for H2 molecule} hydrogen molecule. Each atom section \ref{sec:Dimensional scaling and interpolation for two-electron atoms}--\ref{sec:Dimensional scaling and interpolation for four-electron atoms} has four subsections: A for $D=1$; B for $D\to\infty$; C for $\epsilon_D^{(1)}$, the first-order perturbation terms; D for $\epsilon_3$, the ground-state energy at $D=3$ is obtained from the interpolation formula. For the hydrogen molecule section \ref{sec:Dimensional scaling and interpolation for H2 molecule}, the subsections deal how the internuclear distance $R$ varies in the $D = 1$ and $D \to \infty$ dimensions and mesh into $D = 3$.  Finally, in \ref{sec:Conclusion} we comment on prospects for blending dimensional limits to serve other many-body problems.

\section{\label{sec:Dimensional interpolation formula} Dimensional interpolation}

For dimensional scaling of atoms and molecules the energy erupts to infinity as $D\to 1$ and vanishes as $D\to \infty$. Hence, we adopt scaled units (with hartree atomic units) whereby $E_D = (Z/\beta)^2 \epsilon_D$ and $\beta = \frac{1}{2}(D - 1)$, so the reduced energy $\epsilon_D$ remains finite in both limits. When expressed in a $1/Z$ perturbation expansion, the reduced energy is given by
\begin{equation}
\epsilon_D = -1 +  \epsilon_D^{(1)} \lambda+   \epsilon_D^{(2)} \lambda^2 + ~... ~  \label{perturbation}
\end{equation}
with $\lambda = 1/Z$, where $Z$ is the total nuclear charge of the corresponding atom. The first-order perturbation coefficient is (\ref{perturbation}, \ref{eq4interpolation3dimensionHe}):
\begin{equation}
\epsilon_D^{(1)} = f(D)= \frac{\Gamma(\frac{D}{2}+\frac{1}{2})\Gamma(D+ \frac{1}{2})}{\Gamma(\frac{D}{2})\Gamma(D+1)}. \label{equation2}
\end{equation}
It represents the expectation value, $\langle\frac{1}{r_{12}}\rangle$, of the electron-electron repulsion evaluated with the zeroth-order hydrogenic wave function, $\exp(-r_1-r_2)$. Accordingly, $\epsilon_D^{(1)}$ is universal. For $D = 1, 3, \infty$ the corresponding term $\epsilon_D^{(1)}= 1/2, 5/8, 2^{-1/2}$ respectively.

Our interpolation for atoms, developed in Ref. \citep{Herschbach2017}, weights the dimensional limits by $\delta = 1/D$, providing $\delta \epsilon_1$ and $(1-\delta) \epsilon_\infty$ in a simple analytic formula
\begin{equation}
\epsilon_D = \delta \epsilon_1 + (1-\delta) \epsilon_\infty + [\epsilon_D ^{(1)} - \delta \epsilon_1^{(1)} - (1-\delta) \epsilon_\infty^{(1)}] \lambda , \label{interpolation_formula}
\end{equation}
We aim to illustrate the interpolation formula more fully, presenting results with modest calculations having respectable accuracy for two, three, and four electrons.

For the hydrogen molecule, a different scaling scheme will be used and illustrated. The rescaling of distances is:

\begin{equation}
R \to \delta R^\prime \text{ for } D \to 1 ; R \to (1 - \delta)R^\prime \text{ for } D \to \infty . \label{scalingforinterpolation}
\end{equation}

An approximation for $D = 3$ (where $R = R^\prime$) emerges:

\begin{equation}
\epsilon_3 (R^\prime) = \frac{1}{3} \epsilon_1 ( \frac{1}{3} R^\prime) + \frac{2}{3} \epsilon_\infty ( \frac{2}{3} R^\prime), \label{scaledinterpolationformula}
\end{equation}

on interpolating linearly between the dimensional limits, developed by Loeser in Refs. \cite{frantz1988, Tan_and_Loeser, lopez1993scaling}.

\section{\label{sec:Dimensional scaling and interpolation for two-electron atoms}Two-electrons: Helium}

The formula worked very well for $D=3$, helium with $\lambda = 1/2$:

\begin{equation}
\epsilon_3 = \frac{1}{3} \epsilon_1 + \frac{2}{3} \epsilon_\infty + \left[ \epsilon_3 ^{(1)} - \frac{1}{3} \epsilon_1 ^{(1)} - \frac{2}{3} \epsilon_\infty ^{(1)} \right] \lambda \label{eq4interpolation3dimensionHe}
\end{equation}

The input ingredients are exact limit energies: $\epsilon_1 = -0.788843$ from Ref. \cite{Rosenthal1971}; $\epsilon_\infty = -0.684442$ from Ref. \cite{Herschbach1986}; and the three first-order perturbation terms $\epsilon_D ^{(1)}$ displayed in Eq. (\ref{equation2}). The interpolation delivered $\epsilon_3 = -0.725780$, a result very close to the exact ground-state energy $-0.725931$ \cite{Herschbach1986}. The interpolation accuracy of $2$ millihartrees is better than current density functional theory.

\subsection{\label{subsec:two-electron atom, variational calculations, in one dimension (D=1)} One-dimension: D=1}

We will calculate the ground-state energy of the Hamiltonian operator using the variational principle. It is less accurate than Ref. \cite{Rosenthal1971}, but much easier to deal with two and more electrons \cite{Lapidus1975}. The Hamiltonian with electrons in delta functions is:

\begin{equation}
\mathcal{H} =  -\frac{1}{2} \frac{\partial^2}{\partial r_1^2}  -\frac{1}{2} \frac{\partial^2}{\partial r_2^2}  -  \delta (r_1) -  \delta (r_2) +  \lambda \delta (r_1-r_2) , \label{HamiltonianHe1D}
\end{equation}
with $\lambda = 1/Z$.
The electronic wave function is as follows:
\begin{equation}
\phi (r_1, r_2) = \chi_1(r_1) \chi_2(r_2),
\end{equation}
where the normalized wave functions $\chi_1$ and $\chi_2$ are defined as:
\begin{equation}
\chi_1(r_1) =  (\xi)^{1/2} e^{-{\xi} \mid {r}_1 \mid} \label{chi1heoned}
\end{equation}
and 
\begin{equation}
\chi_2(r_2) =  (\xi)^{1/2} e^{-{\xi} \mid {r}_2 \mid}. \label{chi2heoned}
\end{equation}
 
We optimize the parameter $\xi$, defined in (\ref{chi1heoned}, \ref{chi2heoned}), and calculate the minimum value of the operator $\mathcal{H_\phi} (\xi)$ defined as:
\begin{equation}
\mathcal{H_\phi} (\xi) = \langle \phi \mid  \mathcal{H}  \mid \phi\rangle = \langle \phi \mid  -\frac{1}{2} \frac{\partial^2}{\partial r_1^2}   -\frac{1}{2} \frac{\partial^2}{\partial r_2^2}  -  \delta (r_1) -  \delta (r_2) +  \lambda \delta (r_1-r_2) \mid \phi\rangle. \label{variationalhamiltonian1dexacthe}
\end{equation}

We divide the above Hamiltonian into three parts, where
\begin{equation}
\langle \phi \mid  \mathcal{H}_{KE}  \mid \phi\rangle = \langle \phi \mid   -\frac{1}{2} \frac{\partial^2}{\partial r_1^2}   -\frac{1}{2} \frac{\partial^2}{\partial r_2^2}  \mid \phi\rangle = \xi ^2 
\end{equation}
is the kinetic energy of the two electrons,
\begin{equation}
\langle \phi \mid  \mathcal{H}_{PE}  \mid \phi\rangle = \langle \phi \mid  -  \delta (r_1) -  \delta (r_2) \mid \phi\rangle = -  2\xi
\end{equation}
is the potential energy of the two electrons due to nuclear attraction, and
\begin{equation}
\langle \phi \mid  \mathcal{H}_{ee}  \mid \phi\rangle = \lambda \langle \phi \mid  \delta (r_1-r_2) \mid \phi\rangle = \lambda \frac{\xi}{2}
\end{equation}
is the interaction energy for electron-electron repulsion in the system.\\

We minimize the Hamiltonian operator $\mathcal{H_\phi} (\xi)$ with respect to $\xi$, with
\begin{equation}
\mathcal{H_\phi} (\xi) =  \xi  ^2  -  2\xi  + \lambda \frac{\xi}{2}, \label{equation13}
\end{equation}
such that
\begin{equation}
\frac{d \mathcal{H_\phi}}{d\xi} =  2 \xi  -  2  +  \frac{\lambda}{2} = 0, \label{xihelium}
\end{equation}

and obtain $\xi_0 = 0.875$, which put into Eq (\ref{equation13}) gives the ground-state energy, $\epsilon_1 = -0.765625$. This result is found in Refs. \cite{Stillinger1970,Hylleraas-Pekeris,Lapidus1975}, but it is approximated by $2.9\%$ since noted the exact value is $\epsilon_1 = -0.788843$.

\subsection{\label{subsec:Two-electron atom, variational calculations, in large-D-dimension} Infinite-dimension: $D \to \infty$}

At large-D limit, the effective ground state Hamiltonian for a two electron atom, with inter-electronic correlation can be written as:
\begin{equation}
\mathcal{H} = \frac{1}{2 \sin ^2 \theta} \left(\frac{1}{r_1^2} + \frac{1}{r_2^2}\right) -  \frac{Z}{r_1} -\frac{Z}{r_2} + J (r_1,r_2, \theta) , \label{HamiltonianN=2exact}
\end{equation}
with 
\begin{equation}
J(r_1,r_2, \theta) = \frac{1}{\sqrt{r_1^2 + r_2^2 - 2 r_1 r_2 \cos \theta}}, 
\end{equation}
for an inter-electronic angle $\theta$.

We minimize the above effective-Hamiltonian with respect to the parameters $r_1, r_2$, and $\theta$ respectively, and obtain the corresponding ground state energy to be: $\epsilon_\infty = -0.684442$ (see Table 1 in \cite{Herschbach2017}, and \cite{Hylleraas-Pekeris}).

\subsection{\label{subsec: Two-electron atom and variational calculations in D-dimension} First-order perturbations: $\epsilon_D^{(1)}$}
In two-electron atom, with nuclear charge $Z$, the exact Hamiltonian in $D$-dimension using atomic units can be written as:
\begin{equation}
\mathcal{H} =  -\frac{1}{2} \triangledown _1^2 - \frac{1}{2} \triangledown _2^2 - \frac{1}{r_1} - \frac{1}{r_2} +  \lambda \frac{1}{r_{12}}, \label{HamiltonianHeND}
\end{equation}
where the Laplacian operator $\triangledown_r^2$ in $D$-dimension is defined as:
\begin{equation}
\triangledown_r^2 = \frac{\partial^2}{\partial r^2} + \left(\frac{D-1}{r}\right) \frac{\partial}{\partial r} + \left( \text{angular part involving } \partial_\theta, \partial_\phi \text{etc} \right). \label{LaplacianND}
\end{equation}

For helium-like atoms we consider the two electrons are in a $1s$-like state with spatial part being symmetric (both electrons are in the same state) and the spin part in the antisymmetric spin singlet. The spatial part of the electronic wave function can be written as:
\begin{equation}
\phi (r_1, r_2) = \chi_1(r_1) \chi_2(r_2),
\end{equation}
where the normalized wave functions $\chi_1(r_1)$ and $\chi_2(r_2)$ are defined as:
\begin{equation}
\chi_1(r_1) =  \mathcal{N} e^{-r_1} \label{chi1Nd}
\end{equation}
and 
\begin{equation}
\chi_2(r_2) =  \mathcal{N} e^{-r_2}. \label{chi2Nd}
\end{equation}

The normalization constant $\mathcal{N}$ is calculated as:
\begin{equation}
\mathcal{N} = \frac{2^{D/2}}{\sqrt{(D-1)! \Omega(D)}}, \label{normalizationconstantN}
\end{equation}
with
\begin{equation}
\Omega(D) = \frac{2 \pi^{D/2}}{\Gamma(D/2)} \label{OmegaD}
\end{equation}
is the surface area of an unit sphere in $D$-dimension.

In D-dimension, with the above wave functions, we obtain the following first-order coefficient \cite{Herschbach2017}:
\begin{equation}
\epsilon_D^{(1)} = f(D)= \langle\phi \mid \frac{1}{r_{12}} \mid \phi\rangle =  \frac{\Gamma(\frac{D}{2}+\frac{1}{2})\Gamma(D+ \frac{1}{2})}{\Gamma(\frac{D}{2})\Gamma(D+1)}. \label{ddimfunction}
\end{equation}

As shown in Eq.(\ref{equation2}) and for  $D = 1, 3, \infty$ respectively $\epsilon_D^{(1)} = \frac{1}{2}, \frac{5}{8}, \frac{1}{\sqrt{2}}$.

\subsection{\label{subsec:Interpolation for Helium} Interpolation for D=3}

We use the formula shown in Eq. (\ref{eq4interpolation3dimensionHe}), already noting that the exact limit energies and first-order perturbation terms, gave $\epsilon_3 = -0.725780$; accurate to $0.02\%$. If we replace the variational result $\epsilon_1 = -0.765625$ (from \ref{subsec:two-electron atom, variational calculations, in one dimension (D=1)} subsection), the formula would give $\epsilon_3 = -0.71839$, accurate to $2.9\%$. However, if we evaluate $\epsilon_1$ by using Eq. (\ref{interpolation_formula}), a subformula is
\begin{equation}
\epsilon_D = \epsilon_\infty + \left[ \epsilon_D ^{(1)} - \epsilon_\infty ^{(1)} \right] \lambda,
\end{equation}
with $D = 1$. This yielded a good approximation of $0.11\%$ for $\epsilon_1 = -0.787996$, near the exact $\epsilon_1 = -0.788843$. With this better $\epsilon_1$ we obtain $\epsilon_3 = -0.725496$, with accuracy of $0.06\%$.

In conventional quantum chemistry textbooks treating $D = 3$ helium, the electron-electron interaction, $\langle1/r_{12}\rangle$, is evaluated by first-order perturbation theory. The result is $\epsilon_3 = -0.687529$ with accuracy of $5.29\%$.

\section{\label{sec:Dimensional scaling and interpolation for three-electron atoms}Three-electrons: Lithium}

The ground-state of the lithium atom had been calculated a long ago by using the variational method with complicated wave functions \cite{james1936ground, weiss1961configuration, larsson1969variational}. Here we present the interpolation formula, using the $D=1$ and $D=\infty$ limits and the first-order perturbation terms. For the ground-state of the lithium atom our formula gave $\epsilon_3 = -0.839648$, with approximation $1.04\%$ compared the exact result $\epsilon_3 = -0.830896$ \cite{scherr1962perturbation}.

\subsection{\label{subsec: Three electron $(N=3)$ Atom in one-dimension}One-dimension: D=1}

In a three-electron atom, with nuclear charge $Z$, the exact Hamiltonian in one-dimension using atomic units can be written as:
\begin{equation}
\mathcal{H} =  \sum_{i=1}^3 \left(  -\frac{1}{2} \frac{\partial^2}{\partial r_i^2}  -   \delta (r_i) \right) + \lambda \sum_{i,j=1}^3 \delta (r_i-r_j) , \label{HamiltonianLi1D}
\end{equation}
with $\lambda = 1/Z$.

In lithium atom we consider two electrons are in $1s$ state and third electron is in a $2s$ state, with spatial part being symmetric (both electrons are in the same state) and the spin part in the antisymmetric state. We write spatial part of the electronic wave function as:
\begin{equation}
\phi (r_1, r_2, r_3) = \chi_1(r_1) \chi_2(r_2) \chi_3(r_3).
\end{equation}
 
The two normalized wave functions $\chi_1(r_1)$, $\chi_2(r_2)$ are described in Eqs. (\ref{chi1heoned}) and (\ref{chi2heoned}). We assume that the $1s$ wave functions are orthogonal to the $2s$ wave function:
\begin{equation}
\chi_3(r_3) =  \left(\frac{9 \xi}{20}\right)^{1/2} \left( \frac{2}{3} - \xi \mid r_3 \mid \right) e^{-{\xi} \mid {r}_3 \mid /2}. \label{Lichi3oned}
\end{equation}

We calculate the ground state energy of a three-electron atom using variational principle. We optimize the parameter $\xi$, defined in the wave functions $\chi_1(r_1), \chi_2(r_2), \chi_3(r_3)$, and obtain the minimum value of the Hamiltonian operator $\mathcal{H_\phi} (\xi)$, which is defined as
\begin{equation}
\mathcal{H_\phi} (\xi) = \langle \phi \mid  \mathcal{H}  \mid \phi\rangle = \langle \phi \mid  \sum_{i=1}^3 \left(  -\frac{1}{2} \frac{\partial^2}{\partial r_i^2}  -   \delta (r_i) \right) + \lambda \sum_{i,j=1}^3 \delta (r_i-r_j) \mid \phi\rangle. \label{variationalhamiltonianLi1dexact}
\end{equation}

We divide the above Hamiltonian (\ref{variationalhamiltonianLi1dexact}) into five parts, where
\begin{equation}
\langle \phi \mid  \mathcal{H}_{KE}  \mid \phi\rangle = \langle \phi \mid   \sum_{i=1}^3 -\frac{1}{2} \frac{\partial^2}{\partial r_i^2}   \mid \phi\rangle = \frac{1}{2} \left(2 \xi ^2 + \frac{17}{20} \xi ^2 \right)
\end{equation}
is the kinetic energy of the three electrons,
\begin{equation}
\langle \phi \mid  \mathcal{H}_{PE}  \mid \phi\rangle = \langle \phi \mid  - \sum_{i=1}^3  \delta (r_i) \mid \phi\rangle = -  \left(2 \xi + \frac{\xi}{5}\right)
\end{equation}
is the potential energy of the three electrons due to nuclear attraction, and
\begin{equation}
\langle \phi \mid  \mathcal{H}_{12}  \mid \phi\rangle = \langle \phi \mid \lambda \delta (r_1- r_2) \mid \phi\rangle = \lambda  \frac{\xi}{2} ,
\end{equation}
\begin{equation}
\langle \phi \mid  \mathcal{H}_{13}  \mid \phi\rangle = \lambda \langle \phi \mid  \delta (r_1-r_3) \mid \phi\rangle  = \lambda \frac{\xi}{15} ,
\end{equation}

\begin{equation}
\langle \phi \mid  \mathcal{H}_{23}  \mid \phi\rangle = \lambda \langle \phi \mid  \delta (r_2-r_3) \mid \phi\rangle =  \lambda \frac{\xi}{15} ,
\end{equation}
are the interaction energies for inter-electronic repulsions in the system.\\

We minimize the Hamiltonian operator $\mathcal{H_\phi} (\xi)$ with respect to $\xi$, with
\begin{equation}
\mathcal{H_\phi} (\xi) = \frac{1}{2} \left(2 \xi ^2 + \frac{17}{20} \xi ^2 \right) - \left(2 \xi + \frac{\xi}{5} \right) + \lambda \frac{2 \xi}{15}+ \lambda  \frac{\xi}{2}, \label{hamiltonianonedimforlithium}
\end{equation}

such that
\begin{equation}
\frac{d \mathcal{H_\phi}}{d\xi} =  \frac{57}{20} \xi  -  \frac{11}{5}  +  \frac{19}{30} \lambda = 0, \label{xilithium}
\end{equation}

and obtain $\xi_0 = 0.697856$, which put into Eq (\ref{hamiltonianonedimforlithium}) gives the ground-state energy, $\epsilon_1 = -0.693979$.
 
\subsection{\label{subsec:  Three-electron atom in Large-D limit} Infinite-dimension: $D \to \infty$}

At large-D-limit the effective ground state Hamiltonian for three-electron atoms, with correlation can be written as:
\begin{equation}
\mathcal{H} = \frac{1}{2} \left( \frac{1}{r_1^2} \frac{\Gamma^{(1)}}{\Gamma} + \frac{1}{r_2^2}  \frac{\Gamma^{(2)}}{\Gamma}+ \frac{4}{r_3^2} \frac{\Gamma^{(3)}}{\Gamma} \right) -  \frac{1}{r_1} -\frac{1}{r_2} -\frac{1}{r_3} + \lambda J (r_1,r_2, r_3)  , \label{HamiltonianN=3exact}
\end{equation}
where 
\begin{equation}
J(r_1,r_2, r_3) = \frac{1}{\sqrt{r_1^2 + r_2^2 - 2 r_1 r_2 \gamma_{12}}} + \frac{1}{\sqrt{r_1^2 + r_3^2 - 2 r_1 r_3 \gamma_{13}}} + \frac{1}{\sqrt{r_2^2 + r_3^2 - 2 r_2 r_3 \gamma_{23}}},
\end{equation}
with $\gamma_{ij}= \gamma_{ij}= \cos \theta_{ij}$, and $\theta_{ij}$ is the angle between $r_i$ and $r_j$.  The quantities $\Gamma^{(i)}$ and $\Gamma$ are called the Gramian determinants. In equation (\ref{HamiltonianN=3exact}) the quantity $\frac{\Gamma^{(i)}}{\Gamma}$ is effectively defined as: 
\begin{equation}
\frac{\Gamma^{(i)}}{\Gamma} = 1 + \sum_{\substack{i,j\\ (j\neq i)}} \gamma_{ij} ^2 -  \sum_{\substack{i,j,k\\ (j\neq i\neq k)}} 2 \gamma_{ij} \gamma_{jk} \gamma_{ki}  ~~\text{for} ~ i, j, k = 1, 2, 3 . 
\end{equation}

See page 111, equation (35) in \cite{Zhen1993} for more details.

We minimize the above effective-Hamiltonian with respect to the parameters $r_1, r_2, r_3$, and $\theta_{12} , \theta_{13} , \theta_{23}$ respectively and obtain the corresponding ground state energy $\epsilon _{\infty} = -0.795453$.

\subsection{\label{subsec: Three electron Atom in D-dimension}First-order perturbations: $\epsilon_D^{(1)}$}

As the electrons reside in two orbits, $1s^2 2s$, there are three electron-electron pairs: one $\langle\frac{1}{r_{12}}\rangle$ from $1s^2$, the two others $\langle\frac{1}{r_{13}}\rangle$ and $\langle\frac{1}{r_{23}}\rangle$ from $1s 2s$. Thus each $\epsilon_D^{(1)}$ coefficient is comprised from the three electron pairs:

\begin{equation}
\epsilon_1^{(1)} = 1/2 + 2(1/15) = 0.633333
\end{equation}
\begin{equation}
\epsilon_\infty^{(1)} = 2^{-1/2} + 2 (0.447212) = 1.601531 
\end{equation}
\begin{equation}
\epsilon_3^{(1)} = 5/8 + 2(17/81) = 1.044753
\end{equation}

The $D=1$ item is obtained via subsection \ref{subsec: Three electron $(N=3)$ Atom in one-dimension}. The $D=3$ item is attained from Ref. \cite{wilson1933wave}. Here we will develop both $D=3$ and $D \to \infty$ bringing the third electron akin with the two-electron treatment in subsection \ref{subsec: Two-electron atom and variational calculations in D-dimension}. As the Hamiltonian is evident in equations (\ref{HamiltonianHeND}) and (\ref{LaplacianND}), we start with the electronic wave function:   

\begin{equation}
\phi (r_1, r_2, r_3) = \chi_1(r_1) \chi_2(r_2) \chi_3(r_3).
\end{equation}
The two normalized functions $\chi_1(r_1)$, $\chi_2(r_2)$ are taken care of in Eqs. (\ref{chi1Nd}), (\ref{chi2Nd}), (\ref{normalizationconstantN}) and (\ref{OmegaD}). We assume that the $1s$ wave functions are orthogonal to the $2s$ wave function:
\begin{equation}
\chi_3(r_3) = \mathcal{N}_1 (1- \alpha r_3) e^{- r_3/2}. \label{chi3NdLi}
\end{equation}
The normalization is:
\begin{equation}
\mathcal{N}_1 = \frac{1}{\sqrt{(1+ \alpha^2 D(D+1) - 2 \alpha D)(D-1)!~ \Omega(D)}}, \label{mathcalN1}
\end{equation}
with $\alpha = \frac{3}{2 D}$.

To obtain the first-order terms for $D=3$ and $D \to \infty$ we need to assemble some integrals associated with the key $f(D)$ function shown in Eqs.(\ref{equation2}) and (\ref{ddimfunction}). The output is:
\begin{equation}
\langle\frac{1}{r_{13}}\rangle =  \langle\frac{1}{r_{23}}\rangle = f(D) F\left(\frac{1}{2}, \frac{3-D}{2}; \frac{D}{2}; y \right) \left(\frac{ab}{a+b}\right), \label{r13forli}
\end{equation}
with
\begin{equation}
y = \left( \frac{a-b}{a+b} \right)^2,
\end{equation}
and the hyprgeometric function $F\left(\frac{1}{2}, \frac{3-D}{2}; \frac{D}{2}; y \right)$ enters in (\ref{ddimfunction}).

The parent integral is,  
\begin{eqnarray}
G_D (a,b) &=& \int d^D r_1 \int d^D r_2 \frac{e^{-ar_1}}{r_1} \frac{e^{-br_2}}{r_2} \frac{1}{r_{12}} \nonumber \\ 
&=& N_D F\left(\frac{1}{2}, \frac{3-D}{2}; \frac{D}{2}; y \right) \frac{1}{(ab)^{D-2}(a+b)}, \label{Gab} 
\end{eqnarray}
and
\begin{equation}
N_D = \frac{(4 \pi)^{D-1} \Gamma(D-1) \Gamma(D-\frac{3}{2}) \Gamma(\frac{D-1}{2})^3}{\Gamma(D-1)^2 \Gamma(D/2)}.
\end{equation} 

From $G_D (a, b)$ we compute the following integral:
\begin{eqnarray}
K_D(i,j) &=& \int d^D r_1 \int d^D r_2 ~ e^{-ar_1} e^{-br_2} r_1^{i-1} r_2^{j-1} \frac{1}{r_{12}} \nonumber \\ 
&=& \left(-\frac{\partial}{\partial a}\right)^{i} \left(-\frac{\partial}{\partial b}\right)^{j} G_D (a,b). \label{Kab} 
\end{eqnarray}

In the integrals, we used the normalized wave functions $\chi_1(r_1), \chi_2(r_2)$, and $\chi_3(r_3)$ already specified, such a typical term:

\begin{eqnarray}
\langle\frac{1}{r_{13}}\rangle &&\sim \int d^D r_1 \int d^D r_3 ~ \chi_{1}^\ast(r_1)^\ast \chi_{3}^\ast(r_3)~ \frac{1}{r_{13}}~ \chi_{1}(r_1) \chi_{3}(r_3) \nonumber \\
&&\sim \int d^D r_1 \int d^D r_3 ~ (1- \alpha r_3)^2  e^{-2r_1} e^{-r_3} \frac{1}{r_{13}}. \label{aandbforli}
\end{eqnarray}

From Eq.(\ref{aandbforli}), we see that we have to put $a = 2$ and $b= 1$, so $y = 1/9$. In Eq. (\ref{r13forli}) (55) the hyprgeometric function is available in tabulations \cite{abramowitz1948handbook}. We computed up to $D = 10^6$ to see that the function converges to
\begin{equation}
F\left(\frac{1}{2}, -\frac{D}{2}; \frac{D}{2}; \frac{1}{9} \right) \to 0.948683 ~ \text{for} ~D\to \infty. \label{hypergeometricdtoinfty}
\end{equation}

At the $D\to \infty$ limit
\begin{equation}
\langle\frac{1}{r_{13}}\rangle = \langle\frac{1}{r_{23}}\rangle = (2/3) 2^{-1/2} (0.948683) = 0.447212 .
\end{equation}

For $D = 3$, the function gives
\begin{equation}
F\left(\frac{1}{2}, 0; \frac{3}{2}; \frac{1}{9} \right) = 0.503703
\end{equation}
and
\begin{equation}
\langle\frac{1}{r_{13}}\rangle = \langle\frac{1}{r_{23}}\rangle = (2/3)(5/8) (0.503703) = 17/81.
\end{equation}

\subsection{\label{subsec: Interpolation formula for Lithium atom}Interpolation for D=3}

Again we use the interpolation formula shown in Eq. (\ref{eq4interpolation3dimensionHe}), 
\begin{equation}
\epsilon_3 = \frac{1}{3} \epsilon_1 + \frac{2}{3} \epsilon_\infty + \left[ \epsilon_3 ^{(1)} - \frac{1}{3} \epsilon_1 ^{(1)} - \frac{2}{3} \epsilon_\infty ^{(1)} \right] \lambda , \label{interpolation3dimensionLi}
\end{equation}
now with $\lambda = 1/Z =1/3$. The input from our A, B, C subsections was:
$$
\epsilon_1 = -0.693979, \epsilon_\infty =  -0.795453,
$$
and 
$$
\epsilon_3 ^{(1)} = 1.044753, ~ \epsilon_1 ^{(1)}= 0.633333, ~ \epsilon_\infty ^{(1)} = 1.601531.
$$

Our interpolation gave the Li atom ground-state energy with error $1\%$: 
$\epsilon_3 = −0.839648$, compared with the exact result $\epsilon_3 = −0.830896$  \cite{scherr1962perturbation}.

\section{\label{sec:Dimensional scaling and interpolation for four-electron atoms}Four-electron: Beryllium}

The electronic structure of the beryllium atom is highly interesting because it's implication in different areas of modern science, for e.g. stellar astrophysics and plasmas, high-temperature physics etc. The ground-state energy for the Be-atom has been calculated by applying various methods for e.g. the Configuration Interaction (CI) method with Slater-type orbitals (STOs) \cite{bunge2010configuration}, the Hylleraas method (Hy) \cite{busse1998nonrelativistic}, the Hylleraas-Configuration Interaction method (Hy-CI) \cite{sims2011hylleraas}, and the Exponential Correlated Gaussian (ECG) method \cite{puchalski2013testing, stanke2009five}. In this section we present the dimensional interpolation formula, by using the results from $D=1$ and $D=\infty$ limit, to obtain the ground state energy of the four-electron atoms. With dimensional interpolation we obtain the ground state energy of beryllium atom to be $\epsilon_3 = -0.910325$, compared to the exact energy $\epsilon_3 = - 0.916709$, with a percentage error of 0.6$\%$.

\subsection{\label{subsec: Four electron $(N=4)$ Atom in one-dimension}One-dimension: D=1}

In Four-electron atoms, with nuclear charge $Z = 1/\lambda$, the exact Hamiltonian in one-dimension using atomic units can be written as:
\begin{equation}
\mathcal{H} =  \sum_{i=1}^4 \left(  -\frac{1}{2} \frac{\partial^2}{\partial r_i^2}  -   \delta (r_i) \right) + \lambda \sum_{i,j=1}^4 \delta (r_i-r_j). \label{Hamiltonianbe1D}
\end{equation}

In beryllium atom we consider the first two electrons are in $1s$ states, and other two electrons are in  $2s$ states with spatial part being symmetric (both electrons are in the same state) and the spin part in the antisymmetric state. We write spatial part of the electronic wave function as follows:
\begin{equation}
\phi (r_1, r_2, r_3, r_4) = \chi_1(r_1) \chi_2(r_2) \chi_3(r_3)\chi_4(r_4),
\end{equation}

The three normalized wave functions $\chi_1(r_1)$, $\chi_2(r_2), \chi_3(r_3)$ are described in Eqs. (\ref{chi1heoned}), (\ref{chi2heoned}) and (\ref{Lichi3oned}). We assume that the $1s$ wave functions are orthogonal to the two $2s$ wave functions  $\chi_3(r_3)$ and
\begin{equation}
\chi_4(r_4) =  \left(\frac{9 \xi}{20}\right)^{1/2} \left( \frac{2}{3}  - \xi \mid r_4 \mid \right) e^{-{\xi} \mid {r}_4 \mid /2}. \label{chi4onedBe}
\end{equation}

We calculate the ground state energy of a four-electron atom with variational principle. We optimize the parameter $\xi$, defined in the wave functions $\chi_1(r_1), \chi_2(r_2), \chi_3(r_3), \chi_4(r_4)$, and obtain the minimum value of the Hamiltonian operator $\mathcal{H_\phi} (\xi)$, which is defined as:
\begin{equation}
\mathcal{H_\phi} (\xi) = \langle \phi \mid  \mathcal{H}  \mid \phi\rangle = \langle \phi \mid  \sum_{i=1}^4 \left(  -\frac{1}{2} \frac{\partial^2}{\partial r_i^2}  -   \delta (r_i) \right) +  \lambda \sum_{i,j=1}^4 \delta (r_i-r_j) \mid \phi\rangle. \label{variationalhamiltonianBE1dexact}
\end{equation}

We divide the above Hamiltonian into five  parts, where
\begin{equation}
\langle \phi \mid  \mathcal{H}_{KE} \mid \phi\rangle = \langle \phi \mid   \sum_{i=1}^4 -\frac{1}{2} \frac{\partial^2}{\partial r_i^2}   \mid \phi\rangle = \left(\xi ^2 + \frac{17}{20} \xi^2\right)
\end{equation}
is the kinetic energy of the four electrons,
\begin{equation}
\langle \phi \mid  \mathcal{H}_{PE}  \mid \phi\rangle = \langle \phi \mid  - \sum_{i=1}^4  \delta (r_i) \mid \phi\rangle = - 2 \left(\xi +  \frac{1}{5}\xi\right)
\end{equation}
is the potential energy of the four electrons due to nuclear attraction, and
\begin{equation}
\langle \phi \mid  \mathcal{H}_{12}  \mid \phi\rangle = \langle \phi \mid \lambda \delta (r_1- r_2) \mid \phi\rangle = \lambda \frac{\xi}{2},
\end{equation}

\begin{equation}
\langle \phi \mid  \mathcal{H}_{i3}  \mid \phi\rangle = \langle \phi \mid \lambda \delta (r_i-r_3)  \mid \phi\rangle = \lambda \frac{\xi}{15} = \langle \phi \mid  \mathcal{H}_{i4}  \mid \phi\rangle , ~ \text{ for $i=1,2,$}
\end{equation}

\begin{equation}
\langle \phi \mid  \mathcal{H}_{34}  \mid \phi\rangle = \langle \phi \mid \lambda \delta (r_3-r_4) \mid \phi\rangle  = \lambda \frac{71}{800}\xi
\end{equation}

are the interaction energies for inter-electronic repulsions in the system.\\

We minimize the Hamiltonian operator $\mathcal{H_\phi} (\xi)$ with respect to $\xi$, with

\begin{equation}
\mathcal{H_\phi} (\xi) = \left( \xi ^2 + \frac{17}{20} \xi ^2 \right) - 2 \left( \xi + \frac{\xi}{5} \right) + \lambda \frac{4 \xi}{15}+ \lambda  \frac{\xi}{2} + \lambda \frac{71}{800}\xi, \label{hamiltonianonedimforberyllium}
\end{equation}

such that

\begin{equation}
\frac{d \mathcal{H_\phi}}{d\xi} =  \frac{37}{10} \xi  -  \frac{12}{5}  +  \frac{2053}{2400} \lambda = 0, \label{xiberyllium}
\end{equation}
and obtain $\xi_0 = 0.590850$,  which put into Eq (\ref{hamiltonianonedimforberyllium}) gives the ground-state energy, $\epsilon_1 = -0.645842$.
 
\subsection{\label{subsec:  Four-electron atom in Large-D limit} Infinite-dimension: $D \to \infty$}

In large-D-limit the effective ground state Hamiltonian for four-electron atoms, with inter-electronic correlation can be written as:

\begin{equation}
\mathcal{H} = \frac{1}{2} \left( \frac{1}{r_1^2} \frac{\Gamma^{(1)}}{\Gamma} + \frac{1}{r_2^2}  \frac{\Gamma^{(2)}}{\Gamma}+ \frac{4}{r_3^2} \frac{\Gamma^{(3)}}{\Gamma} + \frac{4}{r_4^2} \frac{\Gamma^{(4)}}{\Gamma} \right) -  \frac{1}{r_1} -\frac{1}{r_2} -\frac{1}{r_3}-\frac{1}{r_4} + \lambda J (r_1,r_2, r_3, r_4), \label{HamiltonianN=3exactbe}
\end{equation}
where 
\begin{equation}
J(r_1,r_2, r_3, r_4) =  \sum_{i,j=1}^4 \frac{1}{\sqrt{r_i^2 + r_j^2 - 2 r_i r_j \gamma_{ij}}},
\end{equation}
with $\gamma_{ij}= \gamma_{ij}= \cos \theta_{ij}$, and $\theta_{ij}$ are the angle between $r_i$ and $r_j$.  The quantities $\Gamma^{(i)}$ and $\Gamma$ are the Gramian determinants. In equation (\ref{HamiltonianN=3exactbe}) the quantity $\frac{\Gamma^{(i)}}{\Gamma}$ is effectively defined as follows: 
 
\begin{equation}
\frac{\Gamma^{(i)}}{\Gamma} = 1 + \sum_{\substack{i,j\\ (j\neq i)}} \gamma_{ij} ^2 -  \sum_{\substack{i,j,k\\ (j\neq i\neq k)}} 2 \gamma_{ij} \gamma_{jk} \gamma_{ki} +\sum_{\substack{i,j,k,l \\ (j\neq i\neq k \neq l)}} \left(2 \gamma_{ij} \gamma_{jk} \gamma_{kl}\gamma_{li}- \gamma_{ij}^2 \gamma_{kl}^2 \right)  ~\text{for} ~ i, j, k,l = 1, 2, 3, 4.
\end{equation}
See page 111, equation (35) in \cite{Zhen1993} for more details.

We minimize the above effective-Hamiltonian with respect to the parameters $r_1, r_2, r_3, r_4$, and $\theta_{12}, \theta_{13}, \theta_{14}, \theta_{23},\theta_{24}, \theta_{34}$ respectively and obtain the corresponding ground state energy $\epsilon_\infty = -0.875837$.

\subsection{\label{subsec: Four electron Atom in D-dimension}First-order perturbations: $\epsilon_D^{(1)}$}

As the electrons reside in two orbits, $1s^2 2s^2$, there are six electron-electron pairs: one $\langle\frac{1}{r_{12}}\rangle$ from $1s^2$, four others $\langle\frac{1}{r_{13}}\rangle$, $\langle\frac{1}{r_{14}}\rangle$, $\langle\frac{1}{r_{23}}\rangle$, $\langle\frac{1}{r_{24}}\rangle$ from $1s 2s$; and another lonely $\langle\frac{1}{r_{34}}\rangle$ from $2s^2$. Each $\epsilon_D^{(1)}$ coefficient is comprised from the six electron pairs:

\begin{equation}
\epsilon_1^{(1)} = 1/2 + 4(1/15) + (71/800) = 0.855417,
\end{equation}
\begin{equation}
\epsilon_\infty^{(1)} = 2^{-1/2} + 4(0.447212) + 0.353553 = 2.849508,
\end{equation}
\begin{equation}
\epsilon_3^{(1)} = 5/8 + 4(17/81) + 0.275696 = 1.740202,
\end{equation}

The $D=1$ item is obtained via subsection \ref{subsec: Four electron $(N=4)$ Atom in one-dimension}. Here we will develop both $D=3$ and $D \to \infty$ bringing the fourth electron akin with the three-electron treatment in subsection \ref{subsec: Three electron Atom in D-dimension}. As the Hamiltonian is evident in equations (\ref{HamiltonianHeND}) and (\ref{LaplacianND}), we start with the electronic wave function:   
\begin{equation}
\phi (r_1, r_2, r_3, r_4) = \chi_1(r_1) \chi_2(r_2) \chi_3(r_3)\chi_4(r_4),
\end{equation}

The two normalized $1s$ wave functions $\chi_1(r_1)$, $\chi_2(r_2)$ are taken care of in Eqs. (\ref{chi1Nd}), (\ref{chi2Nd}), (\ref{normalizationconstantN}) and (\ref{OmegaD}). We assume that the $1s$ wave functions are orthogonal to the $2s$ wave functions $\chi_2(r_2)$, defined in \ref{chi3NdLi}, and :
\begin{equation}
\chi_4(r_4) = \mathcal{N}_1 \left(1- \alpha r_4\right)  e^{- {r_4}/2}, \label{BEchi4Nd}
\end{equation}
with normalization constant $\mathcal{N}_1$ defined in \ref{mathcalN1}.

We take same approach as subsection \ref{subsec: Three electron Atom in D-dimension} to calculate the first-order term (the $2s^2$ electron-electron repulsion term) at $D \to \infty $ limit with the help of equations (\ref{Gab}, \ref{Kab}):
\begin{equation}
\langle\frac{1}{r_{34}}\rangle = f(D) F\left(\frac{1}{2}, \frac{3-D}{2}; \frac{D}{2}; y \right) \left(\frac{ab}{a+b}\right) \label{r34forbe}
\end{equation}
with, $y = \left(\frac{a-b}{a+b}\right)^2$ and $f(D)$ function shown in Eqs.(\ref{equation2}) and (\ref{ddimfunction}). This is same functional expression as in lithium atom (\ref{aandbforli}), but the arguments are different.

To calculate the first-order perturbation coefficient $\langle\frac{1}{r_{34}}\rangle$ for beryllium we use the  normalized wave functions $\chi_{1}(r_1), \chi_{2}(r_2), \chi_{3}(r_3)$ and $\chi_{4}(r_4)$ already specified, which gives  rise to a typical term like
\begin{eqnarray}
\langle\frac{1}{r_{34}}\rangle &&\sim \int d^D r_3 \int d^D r_4 ~ \chi_{3}^\ast(r_3) \chi_{4}^\ast(r_4)~ \frac{1}{r_{34}}~ \chi_{3}(r_3) \chi(r_4) \nonumber \\
&&\sim \int d^D r_3 \int d^D r_4 ~ (1- \alpha r_3)^2(1- \alpha r_4)^2  e^{-r_3} e^{-r_4} \frac{1}{r_{34}}. \label{aandbforbe}
\end{eqnarray}

From the above Eq. (\ref{aandbforbe}), we see that we have to put  $a=1$ and $b=1$ , so $y = 0$. In Eq.(\ref{r34forbe}) the hyprgeometric function
\begin{equation}
\lim_{D\to \infty} F\left(\frac{1}{2}, \frac{3-D}{2}; \frac{D}{2}; y \right) = \lim_{D\to \infty} F\left(\frac{1}{2}, -\frac{D}{2}; \frac{D}{2}; 0 \right) =1, \label{hypergeomdforbe}
\end{equation}
and $ f(D) \to  2^{-1/2}$ at $D\to \infty$ limit.

At $D\to \infty$ limit (\ref{r34forbe}) gives
\begin{equation}
\langle\frac{1}{r_{34}}\rangle = 0.353553.
\end{equation}

For $D=3$ we use the following formula from \cite{Stillinger1975} and \cite{abramowitz1948handbook}:
\begin{eqnarray}
G^k_3 (a,b) &=& \int d^3 r_1 \int d^3 r_2 \frac{e^{-ar_1}}{r_1} \frac{e^{-br_2}}{r_2} r_{12}^{k-1} \nonumber \\ 
&=& (4 \pi)^2 \Gamma(k+1) \left( a^2 - b^2 \right)^{-1} \left( b^{-k-1} - a^{-k-1} \right), \label{G3ab} 
\end{eqnarray}

From the above relation (\ref{G3ab}) we can compute the following integral:
\begin{eqnarray}
K_3(i,j,k) &=& \int d^3 r_1 \int d^3 r_2 ~ e^{-ar_1} e^{-br_2} r_1^{i-1} r_2^{j-1} r_{12}^{k-1} \nonumber \\ 
&=& \left(-\frac{\partial}{\partial a}\right)^{i} \left(-\frac{\partial}{\partial b}\right)^{j} G^k_3 (a,b). \label{K3ab} 
\end{eqnarray}
At $D=3$ the $2s$ wave function 
\begin{equation}
\psi_{2s} (r) =  \sqrt{\frac{\alpha^3}{32 \pi}} (2- \alpha  r)  e^{- \alpha r/2}, \label{chi2threebe}
\end{equation}
with $\alpha =1$ such that
\begin{equation}
\langle\frac{1}{r_{34}}\rangle = \int d^3 r_3 \int d^3 r_4 ~ | \psi_{2s} (r_3) |^2 ~ \left( \frac{1}{r_{34}} \right)~ |\psi_{2s} (r_4)|^2. \label{r34threedimbe} 
\end{equation}
To calculate the inter-electronic repulsion energy $\langle\frac{1}{r_{34}}\rangle$ from (\ref{r34threedimbe}) we use the above type of integrals $G^k_3 (a,b)$ in Eq. (\ref{G3ab}) and $K_3(i,j,k)$ in Eq. (\ref{K3ab}), with $a=1 , ~ b=1,$ and $k=0$. 

With the help of (\ref{G3ab}, \ref{K3ab}) we calculate the first-order coefficient ($2s$-$2s$ part) for beryllium atom in three dimension:
\begin{equation}
\langle\frac{1}{r_{34}}\rangle = 0.275696.
\end{equation}

\subsection{\label{subsec: Interpolation formula for Beryllium atom}Interpolation for D=3}

We again use the interpolation formula shown in Eq. (\ref{eq4interpolation3dimensionHe}), 
\begin{equation}
\epsilon_3 = \frac{1}{3} \epsilon_1 + \frac{2}{3} \epsilon_\infty + \left[ \epsilon_3 ^{(1)} - \frac{1}{3} \epsilon_1 ^{(1)} - \frac{2}{3} \epsilon_\infty ^{(1)} \right] \lambda , \label{interpolation3dimensionLi}
\end{equation}
now with $\lambda = 1/Z =1/4$. The input from our A, B, C subsections was:

$$
\epsilon_1 = -0.645842, \epsilon_\infty =  0.875837,
$$
and 
$$
\epsilon_3 ^{(1)} = 1.740202, ~ \epsilon_1 ^{(1)}= 0.855417, ~ \epsilon_\infty ^{(1)} = 2.849508.
$$

Our interpolation gives the Be atom ground-state energy with error $0.6\%$: 
$\epsilon_3 = -0.910325$, compared with the exact result $\epsilon_3 = - 0.916709$.

\section{\label{sec:Dimensional scaling and interpolation for H2 molecule}Hydrogen molecule}

The ground state potential energy function, $V(R)$, of the hydrogen molecule has been calculated by many methods \cite{Coolidge1933, kolos1960,svidzinsky2005,chen2005}. Recently, Turbiner, et al \cite{olivares2019towards} presented a general theory for obtaining the $V(R)$ function for diatomic molecules. They dealt with the Born-Oppenheimer approximation, based on matching $R$ in short and long distances via a two-point Pad\'e approximation. Here, we present a simpler approach obtaining $V(R)$ for H$_2$ at $D = 3$ by using interpolation between $D = 1$ and $D \to \infty$ dimensional limits. Key aspects of dimensional scaling had been developed years ago by Loeser, et al \cite{Tan_and_Loeser, lopez1993scaling} and Frantz \cite{frantz1988}. They did an excellent treatment on H$_2^+$ and partial on H$_2$. Now we will complete  $V(R)$ for H$_2$ by interpolation.

\subsection{\label{subsec:H2 molecule in one dimension} One-dimension: D=1}

In H$_2$, with nuclear charge of each atom $Z$, the electronic part of the Hamiltonian in one-dimension using atomic units can be written as \cite{Lapidus1975, Lapidus1982}:

\begin{equation}
\mathcal{H} =  -\frac{1}{2} \frac{\partial^2}{\partial r_1^2} -\frac{1}{2} \frac{\partial^2}{\partial r_2^2}  -  \delta (r_1-a) -  \delta (r_1+a) -  \delta (r_2-a) -  \delta (r_2+a) + \lambda \delta (r_1-r_2), \label{Hamiltonianh21D}
\end{equation}

with $a= R/2$, where $R$ is the distance between the two nuclei  located at $ r= \pm a$; also $\lambda = 1/Z = 1$. The Hamiltonian energy eigenvalues provide symmetric and antisymmetric states under exchange of the electrons. The symmetric state pertains to the ground-state potential energy \cite{Lapidus1975}:

\begin{equation}
\epsilon_1 (R) = - \frac{1+\left( 4 + 2 R + R ^2\right) e^{-2R}}{1+\left( 1 + R \right)^2 e^{-2R}}, \label{e1forhydrogen}
\end{equation}

The total binding energy is obtained by adding the nucleus-nucleus-interaction term ($1/R$) with the electronic energy.

\subsection{\label{subsec:H2 molecule at the large-D-dimension limit} Infinite-dimension: $D \to \infty$}

For H$_2$, convention locates the two nuclei A and B on the $z$-axis at $-R/2$ and $R/2$, respectively, with equal charges $Z_A = Z_B = Z$. The electrons  are located at ($\rho_1, z_1$) and ($\rho_2, z_2$), with a dihedral angle $\phi$ specifying their relative azimuthal orientation about the molecular axis. The effective Hamiltonian for large-D limit in cylindrical coordinates is \cite{herrick1975, frantz1988}:

\begin{equation}
\mathcal{H} = \frac{1}{2} \left( \frac{1}{\rho_1^2} + \frac{1}{\rho_2^2} \right) \frac{1}{\sin ^2 \phi} -  \sum_{i=1}^2 \left[ \frac{Z}{\sqrt{\rho_i^2 + (z_i + a)^2}} + \frac{Z}{\sqrt{\rho_i^2 + (z_i - a)^2}} \right] + J (\rho_1,\rho_2, z_1, z_2, \phi), \label{Hamiltonianh2moleculeinf}
\end{equation}

with $a=R/2$ and

$$
J (\rho_1,\rho_2, z_1, z_2, \phi) = \frac{1}{\sqrt{(z_1 - z_2)^2 + \rho_1^2 + \rho_2^2 - 2 \rho_1 \rho_2 \cos \phi}}.
$$

In the $D \to \infty$ limit, the Hamiltonian has two locations for electrons, namely: symmetric, with $\rho_1 = \rho_2$ and $z_1 = z_2$, and antisymmetric, with $\rho_1 = \rho_2$ and $z_1 = -z_2$. When $R$ has the nuclei well apart, in the symmetric case, both electrons cluster near one of the nuclei $(\text{H}_2 \to \text{H}^- + \text{H}^+)$; in the antisymmetric case, each electron resides near just one of the nuclei $(\text{H}_2 \to \text{H} + \text{H})$. Thus, the antisymmetric case is much more favorable for the ground-state energy.

We minimize the Hamiltonian (\ref{Hamiltonianh2moleculeinf}) with respect to $\rho$'s and $z$'s to obtain the ground state energy, $\epsilon_\infty (R)$; we numerically evaluate the corresponding optimized parameters $ \rho_1^*, \rho_2^*, z_1^*, z_2^*$, and $\phi^*$ for different values of $R$.

The total binding energy is obtained by adding to $\epsilon_\infty (R)$  the internuclear-interaction term $(1/R)$.

\subsection{\label{subsec: Interpolation formula for H2 molecule}Interpolation for D=3}

Unlike the atoms, our interpolation will be different for a molecule. An atom has only one nucleus, with the electrons orbiting about the positive charge; then our interpolation deals with the first-order perturbation works well but not for a molecule. For a diatomic molecule, $V(R)$ is fundamental, with R distance roaming between the nuclei. As mentioned in Eqs. (\ref{scalingforinterpolation}) and  (\ref{scaledinterpolationformula}), our interpolation for H$_2$ uses a modified rescaling scheme developed by Loeser \cite{frantz1988, Tan_and_Loeser, lopez1993scaling} with the $D = 1$ and $D \to \infty$ dimensional limits:

\begin{equation}
\epsilon_3 (R) = \frac{1}{3} \epsilon_1 (R) + \frac{2}{3} \epsilon_\infty (R), \label{interpolation3dimensionh2}
\end{equation}

The rescaled distances are:

\begin{subequations}\label{rescaled_distance}
\begin{eqnarray}
\text{In } D = 1: ~ r^\prime_i \to r^\prime_i/3 \text{ and } R \to R^\prime/3 , \text{ for } i = 1, 2  \,;
\end{eqnarray}
\begin{eqnarray}
\text{In } D \to \infty: ~ \rho_i \to 2 \rho^\prime_i/3, z _i \to 2 z^\prime_i/3 ,  \text{ and } R \to 2 R^\prime/3 ,  \text{ for } i = 1, 2   \,.
\end{eqnarray}
\end{subequations}

The rescaled Hamiltonians have distinct factors in the kinetic and potential energy parts:

\begin{subequations}\label{rescaled_Hamiltonian}
In $D = 1$: Hamiltonian (\ref{Hamiltonianh21D}) becomes:

\begin{eqnarray}
\mathcal{H}_{D=1} =  -\frac{9}{2} \frac{\partial^2}{\partial r_1^2} -\frac{9}{2} \frac{\partial^2}{\partial r_2^2}  -  3\delta (r_1-a) -  3\delta (r_1+a) -  3\delta (r_2-a) -  3\delta (r_2+a) + 3 \lambda \delta (r_1-r_2)  \,. \nonumber \\
\end{eqnarray}

In $D \to \infty$: Hamiltonian (\ref{Hamiltonianh2moleculeinf}) becomes:

\begin{eqnarray}
\mathcal{H}_{D=\infty} = \frac{9}{4} \left( \frac{1}{\rho^2 \sin ^2 \phi} \right) -  3 \left[ \frac{Z}{\sqrt{\rho^2 + (z + a)^2}} + \frac{Z}{\sqrt{\rho^2 + (z - a)^2}} \right] + \frac{3}{2}  J (\rho, z, \phi)   \,, 
\end{eqnarray}
\end{subequations}

with $a = R/2$ and $$J (\rho, z, \phi) = \frac{1}{\sqrt{(2z)^2 + 2\rho^2  - 2 \rho^2 \cos \phi}} \,.$$

We minimized these rescaled Hamiltonians (\ref{rescaled_Hamiltonian}) with respect to the rescaled distances (\ref{rescaled_distance}).

\begin{figure}[H]
    \centering
    \begin{tikzpicture}[
 image/.style = {text width=0.8\textwidth, 
                 inner sep=0pt, outer sep=0pt},
node distance = 1mm and 1mm
                        ] 
\node [image] (frame1)
    {\includegraphics[width=\linewidth]{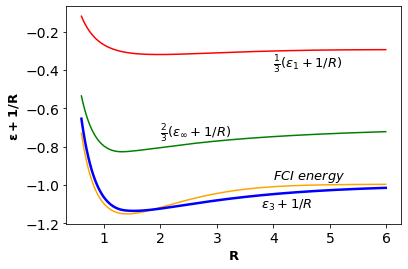}};
\end{tikzpicture}
   \caption{The binding $V(R)$ curves for H$_2$, shown from different dimensions. The red and green curves pertain to $D = 1$ and $D \to \infty$, respectively, parts from Eqs.  (\ref{rescaled_distance}) and  (\ref{rescaled_Hamiltonian}) of the interpolaiton. The blue curve, $\epsilon_3(R) + 1/R$, represents $D = 3$, the interpolation result at Eq. (\ref{interpolation3dimensionh2}). The orange curve is a highly accurate result obtained from computation \cite{mcclean2017openfermion}.} 
    \label{figureh2aftercorrectioncombined}
\end{figure}

In Fig. \ref{figureh2aftercorrectioncombined}, we have plotted the binding energies of H$_2$ as functions of $R$, in the three dimensions (\ref{interpolation3dimensionh2}), adding the nuclear repulsion term, $1/R$. The curves are colored: red for $D = 1$, green for $D \to \infty$, and blue for $D = 3$, the interpolation. It compares fairly well with the nominally exact $V(R)$ curve, colored orange, for H$_2$ obtained from the full configuration interaction (FCI) method \cite{Full_configuration_method, sherrill1999configuration}. We have obtained the FCI by using the OpenFermion quantum computational chemistry software \cite{mcclean2017openfermion}.

\section{\label{sec:Conclusion}Conclusion and prospects}

The formula used for atoms we consider unorthodox, as it recently emerged \citep{Herschbach2017} whereas other $D$-interpolations are elderly \cite{kais1994large, cardy1996scaling}. The fresh aspect links the energies $\epsilon_1$ and $\epsilon_\infty$ together with the first-order perturbation coefficients $\epsilon_1^{(1)}$ and $\epsilon_\infty^{(1)}$ plus $\epsilon_3^{(1)}$ from their $1/Z$ expansions. Those perturbations arise from of electron-electron pair interactions, $\langle1/r_{ij}\rangle$; they actually provide much of the dimension dependence. For H$_2$ we used a different scaling than with the atoms, since H$_2$ links the distance $R$ between the two nuclei. Then the rescaling is: $R \to 1/3 R^\prime$ for $D \to 1$; $R \to 2/3 R^\prime$ for $D \to \infty$. Interpolating between the dimensional limits gave a fair approximation of the binding energy for $D = 3$, when compared with the full configuration interaction (FCI).

In tally, our sections \ref{sec:Dimensional scaling and interpolation for two-electron atoms}  \ref{sec:Dimensional scaling and interpolation for three-electron atoms} \ref{sec:Dimensional scaling and interpolation for four-electron atoms} treat He, Li, Be; in \ref{sec:Dimensional scaling and interpolation for H2 molecule} dealt with H$_2$. In subsections we describe the $D = 1$ limit, the $D = \infty$ limit, the first-order perturbations, and the interpolation output.

The ingredients of the interpolation are well suited for computing. We expect the method to hold true for larger atomic, molecular and extended systems. More than ground-state energies are accessible. However, there are prospects for combining dimensional limits to serve other many-body problems. One is examining dimensional dependence of quantum entanglement \cite{kais2007entanglement, huang2006entanglement}. Another is the isomorphism between the Ising model \cite{chandler1987introduction} and two-level quantum mechanics \cite{tsipis1996new}. Long ago the Ising model was solved in one, two and infinite dimensions \cite{ising1925,Stanley1976,Berlin1952}, as well much activity near four dimensions \cite{herschbach1996dimensional}. The unknown solution at $D = 3$ remains a challenge even by quantum computing \cite{xia2017electronic,Nielsen}. More light on the solution might come by blending of dimensions akin to our unorthodox interpolated formula.

\section*{\label{sec:Acknowledgement}Acknowledgement}

The authors acknowledge the  financial  support by Integrated Data Science Initiative Grant (IDSI F.90000303), Purdue University.

\bibliography{ref}


%
%
%
%
%


\end{document}